# Low Noise Inverse Magnetoelectric Magnetic Field Sensor

L. Thormählen[1,*], P. Hayes[1], E. Elzenheimer[2,3], E. Spetzler[4], J. McCord[4], G. Schmidt[2], M. Höft[3], D. Meyners[1], E. Quandt[1]

[1]Inorganic Functional Materials, Institute for Materials Science, Kiel University, Kiel 24143, Germany
[2]Digital Signal Processing and System Theory, Institute of Electrical Engineering and Information Technology, Kiel University, Kiel 24143, Germany
[3]Microwave Engineering, Institute of Electrical Engineering and Information Technology, Kiel University, Kiel 24143, Germany
[4]Nanoscale Magnetic Materials, Institute for Materials Science, Kiel University, Kiel 24143, Germany
*Correspondence: lath@tf.uni-kiel.de

**Abstract:** In the development of any type of magnetic field sensor based on magnetic films, special consideration must be given to the magnetic layer component. The presented work investigates the use of flux closing magnetostrictive multilayers for inverse magnetoelectric sensors. In such a type of magnetic field sensor, highly sensitive AC and DC field detection relies on strong excitation of the incorporated magnetic layers by piezoelectrically driven cantilever oscillation at mechanical resonances. The provoked periodic flux change is influenced by the magnetic field to be measured and is picked up by a coil, which generates the measured output. The effect of the magnetic multilayer on linearity, noise behavior, and detection limit of DC and AC signals is investigated.
This study demonstrates the next step for inverse magnetoelectric thin film sensors, which achieve one order of magnitude improved detection limits with less than 8 pT/Hz$^{1/2}$ at 10 Hz and 18 pT/Hz$^{1/2}$ at DC using exchange bias stabilized magnetic multilayers for obtaining flux closure.

**Keywords:** Magnetoelectric, piezoelectric, thin film composite, strain modulation, electric frequency conversion, magnetic field sensor, magnetic noise, magnetostrictive multilayers

___________________________________________________________________________________

The wide range of applications for magnetic field sensors is continuing to drive research into magnetic effects such as magnetoresistance in different modifications [1], magnetoimpedance [2], and magnetoelectricity [3]. These effects have been identified as promising for the use in specialized thin film magnetic field sensors characterized by a low Limit-Of-Detection (LOD) at low frequencies [4], a high dynamic range and good linearity [5, 6]. Such characteristics make thin film sensors suitable candidates for automotive [7, 8], navigation [9], and medical applications [10, 11]. Due to the exhibited extremely small magnetic fields, biomagnetism presents a particular challenge for any type of sensor concept. As possible thin film solutions, tunnel magnetoresistance magnetometers [12], giant magnetoresistance magnetometers [13]. and magnetoelectric (ME) composites with strain-mediated coupling have been evaluated [14]. The application capability can be significantly improved by the skillful use of magnetic materials and, especially in the case of film sensors, by tailored magnetic multilayer structures. Custom fabrication of the magnetic layers using standard fabrication methods for Micro-Electro-Mechanical Systems (MEMS) allows magnetic layer property tuning to meet the specific requirements of the sensor concept under consideration [15].

In this work, the inverse magnetoelectric effect is applied as underlying sensor concept [16]: In a thin film composite material designed as cantilever, a mechanical resonance is excited by means of the inverse piezoelectric effect. Via inverse magnetostriction (Villari effect [17]), the magnetization of integrated magnetostrictive layers is modulated accordingly, whereby the external magnetic field (the field to be measured) influences the layers' total magnetization. After inductive detection of the magnetic field dependend magnetization change by a pick-up coil, the signal is demodulated to determine the measurement field. This detection scheme is based on ideas presented by Fetisov *et al.* [16].

Following this approach, we recently examined ME cantilevers with sizes of 24.95 mm × 2.45 mm × 0.35 mm and coated with single layers of amorphous magnetostrictive FeCoSiB (cf. Fig. 1(a)) [4, 18, 19]. The sensors were studied in a specific oscillating mode characterized by bending along the short cantilever axis, i.e., the so-called U-mode, which exhibited a resonance frequency of approximately 500 kHz. To visualize the U-mode deflection of the cantilever in the pick-up coil, a finite element modeling (FEM) result is shown in Figure 1 (b) (s. suppl. mat. for modeling details). Displacement and the connected stress field are inhomogeneously distributed across the cantilever dimension. Additionally, the simulation reveals undesired wave motions occurring at the cantilever edges, which were discussed in more detail by Schmalz *et al.* [20].

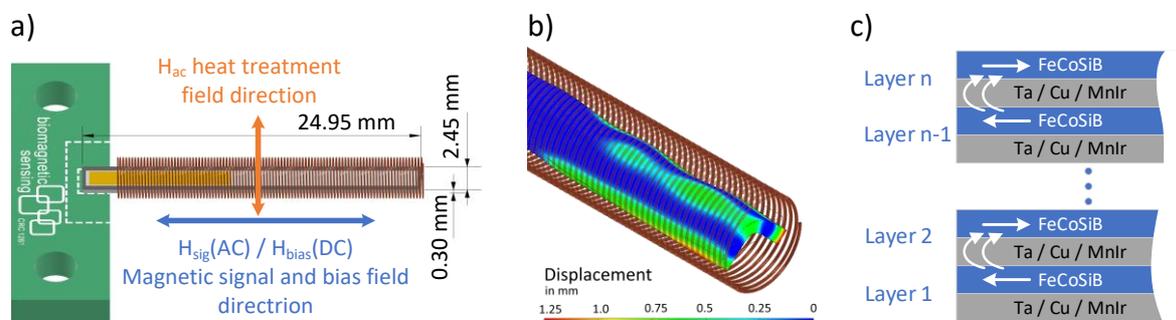

**Figure 1: (a) Schematic of the cantilever sensors used in the inverse magnetoelectric configuration. The illustration shows the pick-up coil covering the entire free length of the cantilever, except at the connection point to the circuit board. For simplification, the electrical contacts and coil connections are not shown. The blue arrow shows the orientation of the measurement and bias fields. The orange arrow indicates the orientation of the magnetic fields applied during annealing. (b) To demonstrate the bending of the cantilever in the U-mode, the cantilever deflection derived from FEM is depicted with a high amplitude magnification. (c) Sketch of advanced layer stack principle for magnetic flux closure. A layer unit is represented by Ta/Cu/MnIr, which is exchange coupled to a FeCoSiB layer. The arrows sketch a possible magnetic flux closure [21] and the preferential direction of magnetization without cantilever excitation.**

With this type of sensor, an LOD of 70 pT/Hz$^{1/2}$ at 10 Hz and 210 pT/Hz$^{1/2}$ at DC and a bandwidth of approximately 1 kHz has been demonstrated [4]. For improvement, consideration of achievable sensor sensitivity and noise is equally important. Noise contributions of magnetic origin can play a dominant role, as has also been shown for fluxgate sensors [22], GMI sensors [23], ΔE effect sensors [24] and magnetoelectric surface acoustic wave (SAW) sensors [25].

In orthogonal fluxgate sensors, the detection performance has been improved by modifying the sensor structure or using DC magnetic fields for sensor biasing and magnetic domain control [26]. More recently, complex magnetic multilayers have been introduced to effectively mitigate magnetic noise in magnetically modulated ME sensors [27]. The layer stack consisted of exchange-biased FeCoSiB layers with layer-wise antiparallel magnetization (Fig. 1(c)). Due to such a flux closure structure, single domain states in the individual layers can be achieved. Consequently, an externally applied magnetic field initiates (mostly) coherent magnetization rotation during the sensor operation, eliminating sources of noise from domain wall activity and reducing hysteretic losses, and as a consequence magnetic noise. At the same time, it has been observed that the stabilization of the magnetic configuration leads to a reduction in the magnetic sensitivity of the sensor due to the reduction from the demagnetization field, which is aligned antiparallel to the exhibited exchange bias. Here, we investigate how the overall performance, determined by sensitivity and noise level, can be improved using multilayers customized for magnetic flux closure in the inverse ME sensor. Compared to the previously investigated antiparallel exchange bias (APEB) layer stacks [27], the thickness of the individual FeCoSiB layers is increased by a factor of five, and a modified heat treatment is applied to reduce the sensitivity reduction of the APEB system and the characteristic noise contributions. We perform magnetic and magnetoelectric characterization of the new inverse sensor. The results demonstrated here are compared with those of a previous inverse single layer sensor [4].

The presented sensors are fabricated by MEMS methods. The piezoelectric layer with a bottom and top electrode of Ta (10 nm)/Pt (100 nm)/AlN (2000 nm)/Cr (10 nm)/Au (100 nm) is sputter-deposited on a double-side polished silicon (350 µm) substrate. The fabrication process has been described in more detail for AlN by Yarar *et al.* [28]. The magnetostrictive multilayer is built up by an 8-fold repetition of Ta (10 nm)/Cu (3 nm)/Mn$_{80}$Ir$_{20}$ (8 nm)/(Fe$_{90}$Co$_{10}$)$_{78}$Si$_{12}$B$_{10}$ (500 nm) (nominal composition in atomic %). A cover layer of Ta (10 nm) is deposited for protection. After patterning the functional layers and dicing the wafer into cantilever structures with dimensions of 24.95 mm × 2.45 mm, the subsequent heat treatment is performed to imprint a magnetic flux closing structure in the cantilever and to introduce a high susceptibility response from the resulting magnetic multilayer structure, by reducing the effective induced magnetic anisotropy contributions. The annealing is conducted at 270 °C in an ambient atmosphere for 30 min, and using a 5 Hz alternating field ($\mu_0 H_{ac,anneal}$) with an amplitude of 20 mT throughout the whole process.

The magnetic properties of the sensors are determined using volumetric magnetic measurement (BH-looper) and magneto-optical Kerr effect (MOKE) microscopy [29]. Volumetric magnetization curves are measured along the longitudinal axis, which represents the sensitivity direction of the ME sensor. For the magnetic domain analysis, the samples were demagnetized along the same axis. For the magnetoelectric sensor analysis, the cantilevers are bonded to one side of the carrier boards. The cantilever is integrated into a pick-up coil, which consists of approximately 800 turns of 100 µm thick enameled copper wire. The connection configuration for the signal amplification and characterization pathway involves the terminals of the carrier board being interfaced with the inputs of a low-noise operational amplifier, specifically the Analog Devices instrumentation amplifier AD8429, configured with a gain set to 10. The amplified output is then directed to a Zurich Instruments HF2LI high-speed lock-in amplifier, which serves as the primary instrument for the subsequent signal characterization process.

During magnetoelectric analysis, the inverse ME sensor is excited by the lock-in amplifier supplying an alternating electric voltage with an RMS amplitude equal to 700 mV at $f_{res}$= 509.25 kHz for the U-mode. The cantilever and the pick-up system are surrounded by a signal and a bias field coil to apply measurement signals as well as optional DC bias fields along the long axis of the sensor (Fig. 1(a)). The setup is positioned in a magnetically and electrically shielded chamber [30]. Both coils are powered by low-noise current sources (Keithley 6221). In DC mode, the magnitude of the induced coil voltage at $f_{res}$ serves as the desired measurement signal. In AC mode, the amplitude spectrum of the sensor output contains sidebands with a frequency separation from $f_{res}$ matching the AC signal field frequency. Due to the periodic piezoelectric driving, the voltage peak at $f_{res}$ appears in the spectrum as a carrier signal (Amplitude Modulation). A small DC bias field is applied to reduce the carrier signal transmission. AC sensor sensitivity, linearity, and equivalent magnetic noise density are investigated.

Multiple field sweeps from −18 to 18 µT are conducted to reveal possible minor hysteresis effects. To demonstrate the achieved DC sensor performance, a periodic DC step measurement is performed for DC bias fields ranging from −4.5 to 4.5 nT in 1 nT steps and the coil signal amplitude was measured for 5 seconds per step. For the measurements of AC magnetic fields, a signal frequency of 10 Hz is used with different amplitudes (1 to 100 nT) and the resulting sideband amplitudes in the sensor spectrum are compared. Additional measurements are conducted at 33 Hz and 70 Hz with a peak-to-peak field amplitude of 2 nT to determine the sensor behavior at different frequencies further.

A comparison of the volumetric magnetization curves and magneto-optical micrographs recorded for a single layer (SL) sensor, as in [4], and the obtained multilayer (ML) sensor is shown in Figure 2. From the magnetization curves shown in Figure 2(a), the SL sample has an approximately two times higher saturation magnetic field in comparison to the ML sample. This reduction in the saturation field is a direct consequence of the alternating field anneal. Another consequence of the increased FeCoSiB layer thickness of 500 nm per layer accompanied by a low unidirectional anisotropy is the multidomain state of the sample visible in Figure 2(b). Although a single domain state can be achieved in multilayer samples, the irregular curved form of the few remaining domains indicates partial antiparallel alignment of magnetization between the layers. More important, no closure domains develop, indicating a dominant stray field coupling between the magnetic layers [21], potentially leading to a reduction of magnetic noise by reducing the domain wall density . Furthermore, the exchange bias pins the magnetic domain walls and thus reduces or eliminates domain wall mobility.

a)

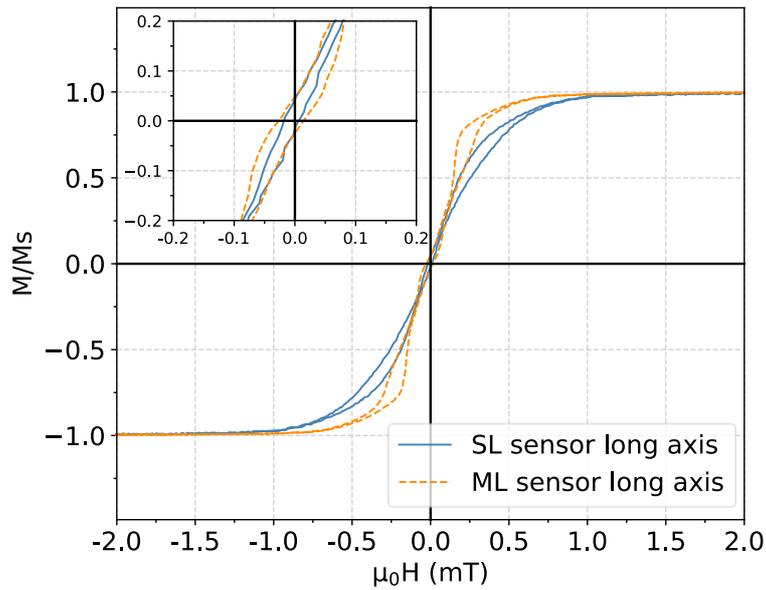

b)

**Domain pattern (single layer)**

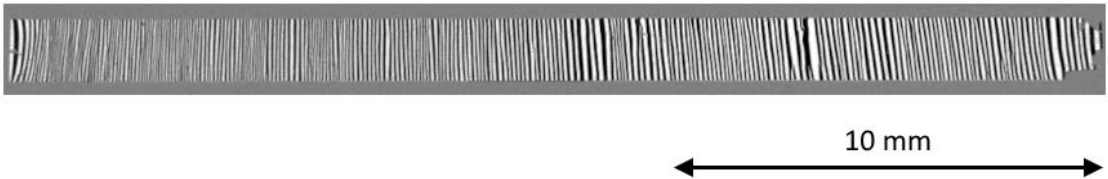

10 mm

**Domain pattern (multilayer)**

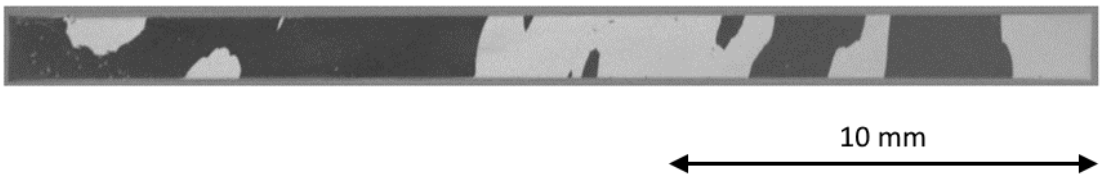

10 mm

**Figure 2**: (a) BH-Looper hysteresis curves comparison of heat-treated SL sample and ML sample (8 x 500 nm exchange bias layers) measured along the long axis of the cantilevers. In both samples the total thickness of FeCoSiB is 4 µm. (b) Magneto-optical Kerr micrographes of the domain patterns of SL and ML samples demagnetized along the long axis of the cantilevers. In both images, the magneto-optical sensitivity axis is oriented along the short side of the cantilevers.

In both cases, areas with different effective anisotropy are visible in the magnetization curves. For all sensors, the anisotropy distribution is influenced by the local stress anisotropy distributions at the edges (Fig. 2(b)). Around zero magnetic field, a small hysteresis opening becomes visible due to individual local switching of magnetic domains or related magnetic domain processes. One reason lies in a magnetic switching at the edges [31], leading to a change of chirality at the magnetic edge curling walls. Another reason is due to individual switching of domains in the different layers with areas of different effective anisotropy in the ML structures. The observed nonlinearity in the magnetization curve constrains the dynamic range of the sensor. This limitation is highlighted in the magnetoelectric characterization shown in Figure 3(a), where the phase-corrected output voltage of the ML sensors is plotted against an applied DC bias field, providing a more accurate representation than considering the absolute voltage would. The zero crossing of the curve is shifted by approx. −3.5 µT due to the magnetic hysteresis and slight misalignment of the unidirectional anisotropy. The slope of the curve (175 ± 3 kV/T between −8 µT and +4 µT) represents the sensor sensitivity to external magnetic DC fields. No hysteresis is visible in the minor loop (±18 µT). However, as noted above, the nonlinearity of the magnetization curve is reflected in the nonlinearity of the sensor sensitivity on the applied bias field.

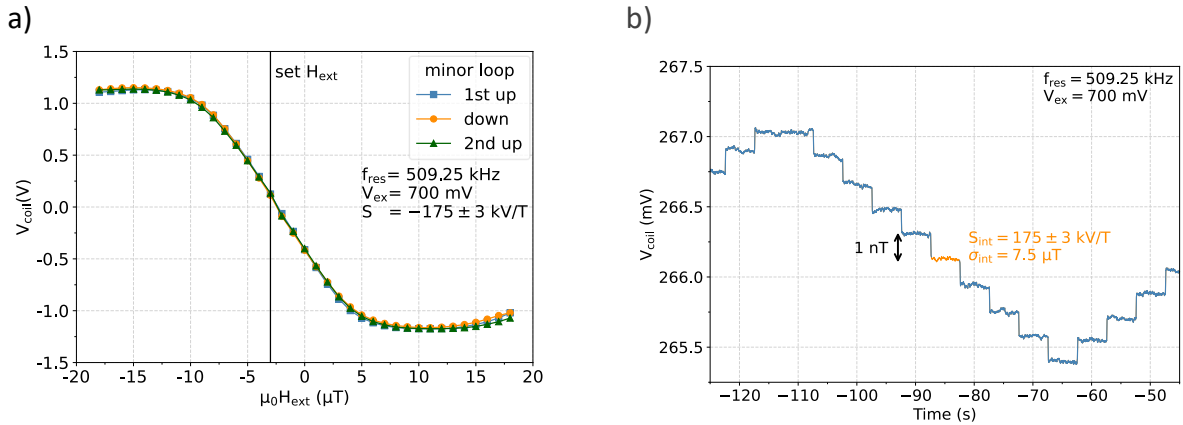

**Figure 3: (a)** Sensor output voltage with unwrapped phase as a function of the applied bias field along the long axis of the cantilever. The measurement was recorded in a loop with magnetic field changing between ±18 µT to estimate the hysteresis of the system. During the measurement the sample was excited at its mechanical resonance at $f_{res}$ = 509.35 kHz. **(b)** Section from the periodic staircase curve relating the coil voltage of the carrier maximum at resonance to the stepwise change of $\mu_0 H_{ext}$ by 1 nT in a range from 4.5 to −4.5 nT over time with 5 s intervals. An exemplary interval of measured data for the DC LOD estimation is marked in yellow.

Figure 3(b) illustrates the periodic staircase approach used to detect possible fluctuations in the system's noise under the influence of DC bias fields. To achieve this, rectangular pulses with a fixed duration of 5 seconds were employed, where the amplitude of each pulse was systematically varied in an alternating pattern, ranging from 4.5 to −4.5 nT over time. Furthermore, the detection limit can be roughly estimated for each field plateau by using the provided sensitivity of 175 ± 3 kV/T and the available noise amplitude spectral density (ASD) with each field step. The ASD within the measurement bandwidth (N) is determined by taking the standard deviation (σ) of the noise data points collected over a given field plateau and dividing it by the Equivalent Noise Bandwidth (ENBW). The ENBW is calculated as the square root of the product of the noise equivalent power (NEP) and the bandwidth of the filter in use. This approach effectively characterizes the noise performance of the system across the specified frequency range. For instance, the yellow data set in Figure 3(b) spanning from -87 to −82 s, exhibiting a σ of 7.5 µV. This calculation process is detailed further in the supplementary material section. Averaging the DC field for all field plateaus computes a low detection limit of 17.5 ±3.4 pT/Hz$^{1/2}$. Compared to the SL sensor properties, this represents a fourfold increase in sensitivity and more than an order of magnitude improvement in the detection limit ($S_{SL}$ = 43 kV/T, detection limit of 210 pT/Hz$^{1/2}$ [4]).

To demonstrate the applicability of the sensor for alternating fields, Figure 4 illustrates coil voltages for the carrier and sideband signals for different applied signal amplitudes (a) and frequencies (c), the sideband amplitude behavior over a large signal field range (b), and the magnetic noise spectral density (d) of the modulated ML sensor. Throughout these measurements, a magnetic bias of $\mu_0 H_{ex} = -3.1\ \mu T$ is applied, which falls into the linear field range of the sensor (Fig. 3(a)) and is motivated by optimal signal-to-noise ratio of the sensor system.

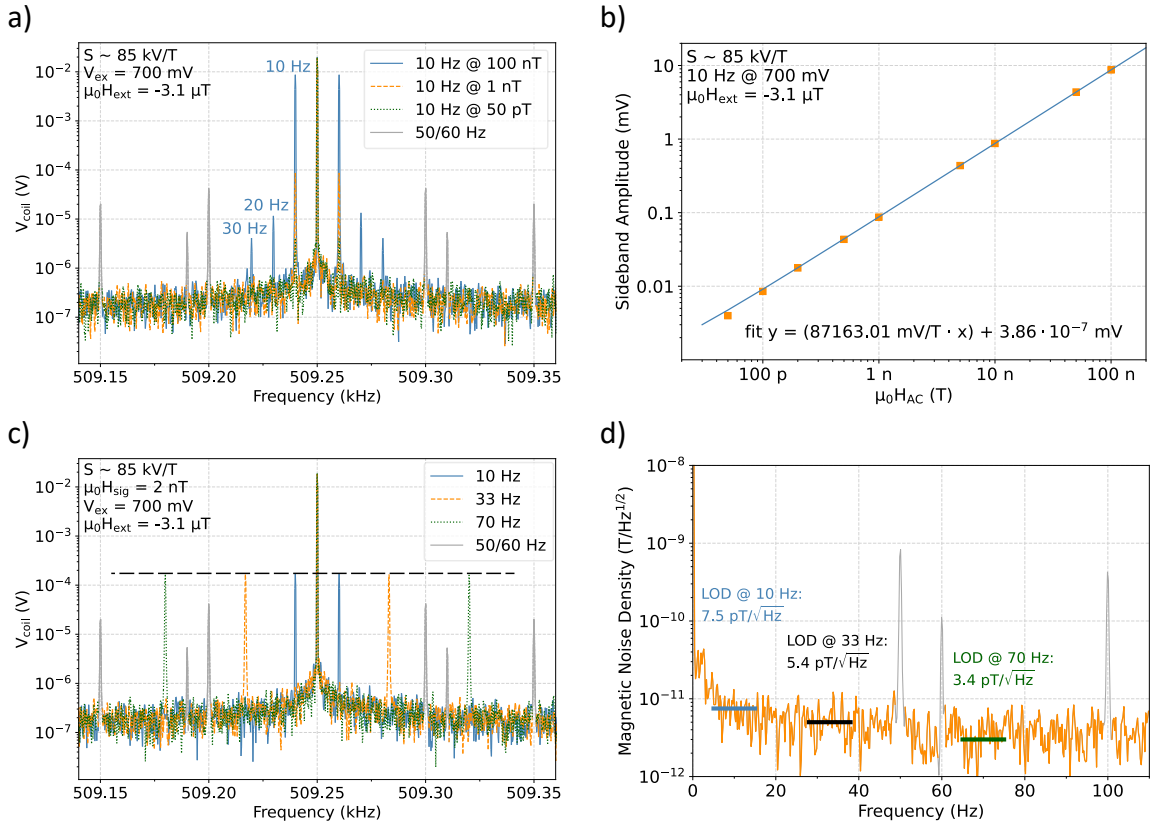

**Figure 4:** (a) Coil voltage spectra recorded with known magnetic test signals of 10 Hz present (50 pT, 1 nT, 100 nT). At large magnetic field ($H_{AC}$) amplitudes, above 100 nT, a slight harmonic distortion of 0.02 % is visible by peaks at Δf = 20 Hz and 30 Hz. For all plots: the gray peaks at $f_{res}$ = ± 50 Hz and ± 60 Hz.

To quantify the linearity, the ML sensor is exposed to 10 Hz signals with various amplitudes from 50 pT to 100 nT (Fig. 4(b)). A linear regression analysis is performed on the sideband peak values, resulting in a residual sum of squares of $3.5 \cdot 10^{-11}$ V² demonstrating a high linear behavior in the investigated measuring range. The signal frequency response of the ML sensor is tested in a range relevant to biomagnetic measurements (10 Hz, 33 Hz and 70 Hz) and for a peak-to-peak signal amplitude of 2 nT. A voltage amplitude of about 167 ±5 µV is measured for all frequencies in the spectral domain (Fig. 4(c)). However, the voltage noise level increases with decreasing frequency distance from the carrier at 509.25 kHz, so the detection limit is frequency dependent. The equivalent magnetic noise density spectrum is displayed in more detail in Figure 4(d). A decrease in the noise floor from DC to 110 Hz is apparent; at 10 Hz, the detection limit is determined to be 7.5 pT/Hz$^{1/2}$ (blue line), at 33 Hz to 5.4 pT/Hz$^{1/2}$ (black line) and 70 Hz to 3.4 pT/Hz$^{1/2}$ (green line).

The results demonstrate that annealing in the AC field of the exchange-biased stack with thick (500 nm) FeCoSiB layers promotes an increased susceptibility (Fig. 2(a)), leading to a high DC sensitivity of approximately 180 kV/T (Fig. 3(a)). In comparison to SL based sensors, magnetic domain density has been significantly reduced, accompanied by a decreased hysteresis within the linear range of ca. 12 µT (Fig. 3a). In addition, the sensor's output noise is reduced by almost an order of magnitude from 3 µV/Hz$^{1/2}$ (SL) [5] to 0.8 µV/Hz$^{1/2}$.

By the application of the magnetic bias $\mu_0 H_{ex}$= −3.1 µT, carrier suppression was employed in all depicted measurements, enabling optimal utilization of the A/D converter's dynamic range for the sensor signals (see supplemental material). This allows the lock-in amplifier to operate in a small amplitude range and to perform the A/D conversation with the available number of bits (14 bits). Consequently, this refinement leads to a lower quantization noise of the digitized signal component and leads to an improvement in the overall sensor performance without changing the sensitivity. However, since there is no signal modulation, accurate sensor calibration is required to effectively account for carrier changes to perform DC field measurements in large field (µT) ranges.

In summary, this work showcases the implementation of an exchange bias system with magnetic flux closure structure for inverse ME sensors, leading to a significant enhancement in sensor performance. This advancement is evidenced by an improvement of one order of magnitude in detection capabilities, achieving a DC detection limit of 18 pT/Hz$^{1/2}$ and an AC detection limit of 7.5 pT/Hz$^{1/2}$ at a frequency of 10 Hz. The significant decrease in domain wall density and simultaneous increase in magnetic susceptibility relative to prior single layer sensors has enhanced sensitivity and lowered the noise floor around to the carrier, paving the way for novel applications. A prime instance is its potential in magnetocardiography (MCG) for R-Peak detection, as evidenced by a promising field study in magnetic cardiology [14].

**Author Contributions:**
Conceptualization, H.P., E.E., L.T., E.S., J.M., D.M, and E.Q.; methodology, H.P., L.T., E.E. and E.S.; validation, H.P., E.E. and L.T.; formal analysis, P.H., E.E., L.T.,D.M. and J.M.; investigation, H.P., E.E., L.T., E.S., and J.M.; resources, H.P., E.E., L.T., and E.S.; data curation, L.T.; writing—original draft preparation, E.E., L.T., E.S., J.M. and D.M.; writing—review and editing, H.P., D.M., J.M. and E.Q.; visualization, L.T.; supervision, D.M. and E.Q.; project administration, D.M. and E.Q.; funding acquisition, D.M. and E.Q.

**Funding:** This research was funded by the German Research Foundation (Deutsche Forschungsgemeinschaft, DFG) through the project A7 and A1 of the Collaborative Research Centre CRC 1261 'Magnetoelectric Sensors: From Composite Materials to Biomagnetic Diagnostics'.

**Data Availability Statement:** The data presented in this study are available on request from the corresponding author.

**Acknowledgments:** The authors would like to thank DFG for funding.

**Conflicts of Interest:** The authors declare no conflict of interest. The funders had no role in the design of the study; in the collection, analyses, or interpretation of data; in the writing of the manuscript, or in the decision to publish the results.

# Supplementary material

# Low Noise Inverse Magnetoelectric Magnetic Field Sensor

L. Thormählen[1], P. Hayes[1], E. Elzenheimer[2,3], E. Spetzler[4], J. McCord[4], G. Schmidt[2], M. Höft[3], D. Meyners[1], E. Quandt[1]

[1]Inorganic Functional Materials, Institute for Materials Science, Kiel University, Kiel 24143, Germany
[2]Digital Signal Processing and System Theory, Institute of Electrical Engineering and Information Technology, Kiel University, Kiel 24143, Germany
[3]Microwave Engineering, Institute of Electrical Engineering and Information Technology, Kiel University, Kiel 24143, Germany
[4]Nanoscale Magnetic Materials, Institute for Materials Science, Kiel University, Kiel 24143, Germany

The simulation in Figure 1(b) was performed using Inventor Professional 2023 64-Bit-Edition, Build: 158, Release: 2023. For the simulation, a cantilever of the dimensions described in the publication was drawn and defined with silicon [100]. The Young's modulus is 131 GPa, and the Poisson number is 0.22. The short section of the cantilever on the side of the circuit board was defined as fixed in the software and a modal analysis was performed. The frequency ranges from 400 to 550 kHz have been simulated, and the resulting displacements were calculated. The resulting U1 mode was then plotted and displayed. The estimation of the detection limit, as detailed above, is predicated upon a Noise-to-Signal Ratio (NSR) analysis. This analysis is conducted using the empirical data presented in Figure 3(b). The results from this figure are instrumental in quantifying the overall system's sensitivity and its capacity to discern magnetic signals from the background noise, thereby providing a measure of the detection limit for the sensor. To estimate a detection limit, the equivalent noise bandwidth (ENBW) needs to be determined. The calculation of the ENBW (suppl. Mat. Eq. 1) involves the square root of the product of the noise equivalent power (NEP = 1.13) and the -3dB filter bandwidth ($B_{-3dB}$ = 7 Hz, 4$^{th}$ Order) [36, pp. 332-333]. In the following, the noise standard deviation (σ) of the respective value range is determined (Fig. 3(b), yellow data area).

$$\mathbf{N} = \frac{\sigma}{\sqrt{\mathbf{NEP} \cdot f_{\text{filter}}}} = \frac{7.5\ \mu V}{\sqrt{1.13 \cdot 7\ \text{Hz}}} \qquad [1]$$

To determine the LOD ENBW (suppl. Mat. Eq. 2), the ENBW is divided by the provided sensitivity (S) determined from the slope of the resulting curves:

$$\mathbf{LOD_{DC}} \sim \frac{\mathbf{N}}{\mathbf{S}} = \frac{2.7 \frac{\mu V}{\text{Hz}^{1/2}}}{171 \frac{\text{kV}}{\text{T}}} \approx 16 \frac{\text{pT}}{\text{Hz}^{1/2}} \qquad [2]$$

As indicated in the main part, using a DC bias field is a way to affect the carrier signal. By minimizing the carrier amplitude to a greater extent, the amplitude disparity between the carrier and desired signal component can be optimized. This can lead to the desired carrier suppression, thereby significantly enhancing the dynamic range of the A/D converter and ultimately reducing the quantization noise.

A field sweep of ±125.5 µT was performed to determine possible carrier signal minima, followed by a measurement with a 1 nT signal with the bias source on and off to observe carrier signal changes (Sup. Fig. 1(a)). An ultralow noise Lake Shore MM81 power supply with a BSC-10 unit was utilized to drive the DC field coil. This setup is indicative of a meticulous approach to minimizing noise, thereby ensuring that the DC field is as stable and clean as possible for precise experimental conditions. The method for estimating the minima carrier amplitude has altered the sensor's magnetic initialization state compared to the previously described measurement. Since a standard initialization procedure or conditioning procedure is not yet established for regular sensor usage, this change is expected to cause variations in magnetic behavior, potentially leading to shifts in the carrier amplitude minima. Three minima occur in the measurement range at $\mu_0 H_{ex}$= -70.4 µT, -16.3 µT and 64.6 µT. For further measurement of the spectral densities (Supplementary Sup. Fig. 1(b)), the magnetic bias is set to 70.4 µT, which leads to a carrier reduction of approximately 600 mV/Hz$^{1/2}$ without influencing the sideband amplitude. In addition, there is a slight reduction in the noise level rise near the carrier signal.

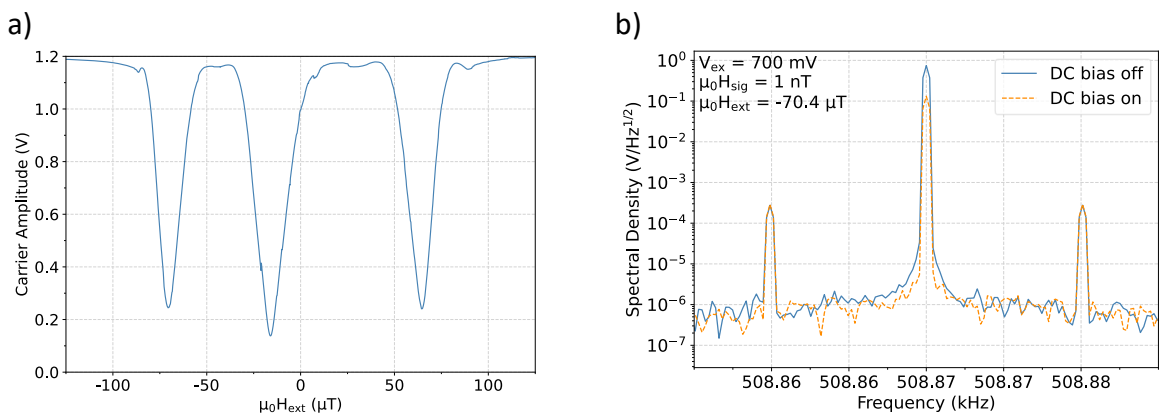

**Supplementary Figure 1 (a) The measured sweep of the magnetic bias field from ± 125 µT as a function of carrier amplitude shows three possible minima for carrier suppression. (b) shows the resulting spectra with the resulting carrier signals. By applying a matched DC bias, the carrier signal can be reduced by 617.64 mV (RMS). This has a positive effect on noise reduction directly at the carrier edges.**